\documentclass[aps,prd, 
twocolumn, showpacs,
superscriptaddress,groupedaddress
,amsmath,amssymb
,nofootinbib
]{revtex4} 

\usepackage{graphicx}
\usepackage{dcolumn}
\usepackage{bm}
\usepackage[usenames,dvipsnames]{color}
\usepackage{amsmath,amssymb}
\usepackage{slashed}
\usepackage{hyperref}
\usepackage{float}
\usepackage[caption = false]{subfig}

\begin{document}

%

\let\a=\alpha      \let\b=\beta       \let\c=\chi        \let\d=\delta
\let\e=\varepsilon \let\f=\varphi     \let\g=\gamma      \let\h=\eta
\let\k=\kappa      \let\l=\lambda     \let\m=\mu
\let\o=\omega      \let\r=\varrho     \let\s=\sigma
\let\t=\tau        \let\th=\vartheta  \let\y=\upsilon    \let\x=\xi
\let\z=\zeta       \let\io=\iota      \let\vp=\varpi     \let\ro=\rho
\let\ph=\phi       \let\ep=\epsilon   \let\te=\theta
\let\n=\nu
\let\D=\Delta   \let\F=\Phi    \let\G=\Gamma  \let\L=\Lambda
\let\O=\Omega   \let\P=\Pi     \let\Ps=\Psi   \let\Si=\Sigma
\let\Th=\Theta  \let\X=\Xi     \let\Y=\Upsilon
%


\def\cA{{\cal A}}                \def\cB{{\cal B}}
\def\cC{{\cal C}}                \def\cD{{\cal D}}
\def\cE{{\cal E}}                \def\cF{{\cal F}}
\def\cG{{\cal G}}                \def\cH{{\cal H}}
\def\cI{{\cal I}}                \def\cJ{{\cal J}}
\def\cK{{\cal K}}                \def\cL{{\cal L}}
\def\cM{{\cal M}}                \def\cN{{\cal N}}
\def\cO{{\cal O}}                \def\cP{{\cal P}}
\def\cQ{{\cal Q}}                \def\cR{{\cal R}}
\def\cS{{\cal S}}                \def\cT{{\cal T}}
\def\cU{{\cal U}}                \def\cV{{\cal V}}
\def\cW{{\cal W}}                \def\cX{{\cal X}}
\def\cY{{\cal Y}}                \def\cZ{{\cal Z}}

%

\newcommand{\Ns}{N\hspace{-4.7mm}\not\hspace{2.7mm}}
\newcommand{\qs}{q\hspace{-3.7mm}\not\hspace{3.4mm}}
\newcommand{\ps}{p\hspace{-3.3mm}\not\hspace{1.2mm}}
\newcommand{\ks}{k\hspace{-3.3mm}\not\hspace{1.2mm}}
\newcommand{\des}{\partial\hspace{-4.mm}\not\hspace{2.5mm}}
\newcommand{\desco}{D\hspace{-4mm}\not\hspace{2mm}}


\def\be{\begin{equation}}
\def\ee{\end{equation}}
\def\bea{\begin{eqnarray}}
\def\eea{\end{eqnarray}}
\def\bm{\begin{matrix}}
\def\em{\end{matrix}}
\def\bpm{\begin{pmatrix}}
	\def\epm{\end{pmatrix}}

{\newcommand{\lsim}{\mbox{\raisebox{-.6ex}{~$\stackrel{<}{\sim}$~}}}
{\newcommand{\gsim}{\mbox{\raisebox{-.6ex}{~$\stackrel{>}{\sim}$~}}}
\def\mpl{M_{\rm {Pl}}}
\def\gev{{\rm \,Ge\kern-0.125em V}}
\def\tev{{\rm \,Te\kern-0.125em V}}
\def\mev{{\rm \,Me\kern-0.125em V}}
\def\ev{\,{\rm eV}}

\title{\boldmath Falsifying leptogenesis for a TeV scale $W^{\pm}_{R}$ at the LHC  }
\author{Mansi Dhuria}
\email{mansi@prl.res.in}
\address{ Physical Research Laboratory, Navrangpura, Ahmedabad 380 009, India.}
\author{Chandan Hati}
\email{chandan@prl.res.in}
\address{ Physical Research Laboratory, Navrangpura, Ahmedabad 380 009, India.}
\address{Indian Institute of Technology Gandhinagar, Chandkheda, Ahmedabad 382 424, India.}
\author{Raghavan Rangarajan}
\email{raghavan@prl.res.in}
\address{ Physical Research Laboratory, Navrangpura, Ahmedabad 380 009, India.}
\author{Utpal Sarkar}
\email{utpal@prl.res.in}
\address{ Physical Research Laboratory, Navrangpura, Ahmedabad 380 009, India.}
\date{\today}

\begin{abstract}
We point out that the discovery of a right-handed charged gauge boson $W_R^\pm$ with mass of around a few TeV, for example through a signal of two leptons and two jets that has been reported by CMS to have a 2.8$\sigma$ local excess or through a signal of a resonance decaying into a pair of standard model (SM) gauge bosons showing a local excess of 3.4$\sigma$ (2.5$\sigma$ global) reported by ATLAS search, will rule out all possibilities of leptogenesis in all classes of the left-right symmetric extensions of the Standard Model (LRSM) with both triplet and doublet Higgs scalars due to the unavoidable fast gauge mediated $B-L$ violating interactions $e_{R}^{\pm} W_{R}^{\mp}  \rightarrow   e_{R}^{\mp} W_{R}^{\pm}$. Our conclusions are very general in the sense that they do not necessarily demand for a lepton number violating detection signal of $W_R^\pm$.

\end{abstract}

\maketitle

  The Left-Right Symmetric Model (LRSM) \cite{Pati:1974yy} is one of the most popular candidates for extensions of the Standard Model (SM) of particle physics. In LRSM the Standard Model gauge group is extended at higher energies to  $${\cal G}_{LR} \equiv SU(3)_c \times SU(2)_{L} \times SU(2)_{R}  \times U(1)_{B-L}$$  where $B-L$ is the difference between baryon (B) and lepton (L) numbers. Left-right symmetry breaking predicts the existence of a massive right-handed charged gauge boson $(W_R^\pm)$. In this Letter, we point out that if $W_R^\pm$ has a mass of a few TeV and can be detected at the LHC, it will have profound consequences for our understanding of the baryon asymmetry of the Universe. This is a unique situation where by observing $W_R^\pm$ at the LHC, we can make a very strong statement about our origin, that is regarding the baryon asymmetry of the Universe. One of the most attractive mechanisms to generate the baryon asymmetry is leptogenesis, in which a lepton asymmetry is created before the electroweak phase transition, which then gets converted to the baryon asymmetry in the presence of ($B+L$) violating anomalous processes \cite{Fukugita:1986hr}. Detection of a $\tev$ scale $W_R^\pm$ at the LHC would imply violation of $(B-L)$ at a lower energy, which will rule out all scenarios of leptogenesis. In this context we must mention that an excess of 2.8 $\sigma$ level was observed in the energy bin $1.8 \tev <M_{lljj} <2.1 \tev$ in the two leptons two jets channel at the LHC by the CMS experiment \cite{Khachatryan:2014dka}, which can be interpreted as due to $W^{\pm}_{R}$ decay by embedding the conventional LRSM with $g_L\ne g_R$ in $SO(10)$ \cite{Deppisch:2014qpa} and with $g_L= g_R$ by taking into account the CP phases and non-degenerate masses of heavy neutrinos \cite{Gluza:2015goa}. More recently, the ATLAS search has also reported a resonance that decays to a pair of SM gauge bosons to show a local excess of 3.4$\sigma$ (2.5$\sigma$ global) in the $W Z$ final state at approximately $2 \tev$ \cite{Aad:2015owa}, which can naturally be explained by a $W_{R}$ in the LRSM framework with a coupling $g_{R}\sim0.4$ \cite{Brehmer:2015cia}.\\

In the LRSM  the fermion sector transforms under the gauge group ${\cal G}_{LR}$  as:
\bea{\label{2.1}}
l_{L} : (1,2,1,-1),\; \;  l_{R} : (1,1,2,-1), \nonumber\\
Q_{L} : (3,2,1,\frac{1}{3}),\; \;  Q_{R} : (3,1,2,\frac{1}{3}).
\eea
In a popular version of the LRSM,
the Higgs sector consists of one bi-doublet $\Phi$ and two triplet $\Delta_{L,R}$ complex scalar fields with the transformations
\be{\label{2.2}}
\Phi : (1,2,2,0),\; \; \Delta_{L} : (1,3,1,2),\; \; \Delta_{R} : (1,1,3,2)
\ee
The left-right symmetry is spontaneously broken to reproduce the Standard Model and the smallness of the neutrino masses can be taken care of by the see-saw mechanism \cite{Minkowski:1977sc}. The symmetry breaking pattern follows the scheme
\bea{\label{2.3}}
{\cal G}_{LR}&&\xrightarrow{\langle \Delta_{R} \rangle} SU(3)_{C}\times SU(2)_{L}\times U(1)_{Y} \equiv {\cal G}_{SM}\nonumber\\
&&\xrightarrow{\langle \Phi \rangle}SU(3)_{C} \times U(1)_{EM}
\equiv {\cal G}_{EM}
\eea
In the first stage of symmetry breaking the right-handed triplet $\Delta_{R}$ acquires a Vacuum Expectation Value (VEV) $\langle \Delta_{R}\rangle=\frac{1}{\sqrt 2}v_{R}$ which breaks the $SU(2)_{R}$ symmetry and gives masses to the $W^{\pm}_{R}$, $Z_{R}$ bosons. The electroweak symmetry is broken by the bi-doublet Higgs $\Phi$, which gives masses to the charged fermions and the gauge bosons $W^{\pm}_{L}$ and $Z_{L}$. The $\Delta_{L}$ gets an induced seesaw tiny VEV, which can give a Majorana mass to the left-handed neutrinos. The generators of the broken gauge groups are then related to the electric charge by the modified Gell-Mann-Nishijima formula
$Q = T_{3L}+T_{3R}+\frac{B-L}{2}$.

In a variant of the LRSM with triplet Higgs scalars, one considers only doublet Higgs scalars to break all the symmetries. This scenario is more popular in all superstring inspired models. Here the Higgs sector consists of doublet scalars
\be {\label{2.9}}
\Phi: (1,2,2,0), \; \; H_{L}: (1,2,1,1), \; \; H_{R}: (1,1,2,1), 
\ee
and there is one additional singlet fermion field $S_{R}$ (1,1,1,0) in addition to the fermions mentioned in Eq. (\ref{2.1}). The doublet Higgs scalar $H_{R}$ acquires a VEV to break the left-right symmetry and results in mixing of $S$ with right-handed neutrinos, giving rise to one light Majorana neutrino, and one heavy pseudo-Dirac neutrino or two Majorana neutrinos.

In the conventional LRSM, the left-right symmetry is broken at a fairly high scale, $M_R > 10^{10} \gev$. First, the gauge coupling
unification requires this scale to be high, and second, thermal leptogenesis in this scenario gives a comparable bound. One often introduces a parity odd scalar and gives a large VEV to this field. This is called D-parity breaking, which may then allow 
$g_L \neq g_R$ even before the left-right symmetry breaking, and hence, this allows gauge coupling unification with TeV
scale $M_R$. This is true for both triplet and doublet models of LRSM. Embedding the LRSM in an $SO(10)$ GUT framework, the violation of D-parity at a high scale can explain the CMS $\tev$ scale $W_{R}$ signal for $g_{R}\approx 0.6 g_{L}$  \cite{Deppisch:2014qpa}. 

For a $\tev$ scale $W^{\pm}_{R}$, all leptogenesis models may be classified into two groups:
\begin{itemize}
\item A lepton asymmetry is generated at a very high scale either in the context of D-parity 
breaking LRSM or through some other interactions, both thermal and non-thermal. 
\item A lepton asymmetry is generated at the TeV scale with resonant enhancement, when the left-right
symmetry breaking phase transition is taking place. 
\end{itemize}
These discussions are valid for the LRSM with both triplet as well as doublet Higgs scalars. We use the reference of the two variants of the LRSM mentioned above to study the lepton number violating washout processes and demonstrate that all these possible scenarios of leptogenesis are falsifiable for a $\tev$ scale $W_{R}$. In models with high-scale leptogenesis with $T > 10^9$~GeV, the low energy $B-L$ breaking is associated with giving mass to the $W_R^\pm$, which allows gauge interactions that wash out all the baryon asymmetry before the electroweak phase transition is over. On the other hand, the same lepton number violating gauge interactions will slow down the generation of the lepton asymmetry for resonant leptogenesis at the TeV scale, so that generation of the required baryon asymmetry of the universe is not possible for TeV scale $W_R^\pm$. 

  The most stringent constraints on the $W^{\pm}_{R}$ mass for  successful high-scale leptogenesis for a hierarchical neutrino mass spectrum ($M_{N_{3R}}\gg M_{N_{2R}}\gg M_{N_{1R}}=m_{N_{R}}$) come from the $SU(2)_{R}$ interactions \cite{Ma:1998sq}. To have successful leptogenesis in the case $M_{N_{R}}>M_{W_{R}}$ the out-of-equilibrium condition for the scattering process  $e_{R}^{-}+W_{R}^{+}\rightarrow N_{R} \rightarrow e_{R}^{+}+W_{R}^{-}$ gives
 \be
 M_{N_{R}}\gsim 10^{16} \gev
 \ee
  with $m_{W_{R}}/m_{N_{R}}\gtrsim 0.1$. Now for the case $M_{W_{R}}>M_{N_{R}} $ leptogenesis can happen either at $T\simeq M_{N_{R}}$ or at $T>M_{W_{R}}$ but at less than the $B-L$ breaking scale.  Considering the out-of equilibrium condition for the scattering process $e_{R}^{\pm} e_{R}^{\pm} \rightarrow W_{R}^{\pm} W_{R}^{\pm}$ through $N_{R}$ exchange one obtains the constraint
  \be
  M_{W_{R}}\gtrsim 3\times 10^{6} \gev (M_{N_{R}}/10^{2} \gev)^{2/3}.
  \ee
   Thus observing a $W_{R}$ signal with a mass in the $\tev$ range for hierarchical neutrino masses rules out the high-scale leptogenesis scenario. In Refs. \cite{Deppisch:2013jxa}, the constraints obtained from the observation of lepton number violating processes and neutrinoless double beta decay were studied to rule out typical scenarios of high-scale thermal leptogenesis, particularly leptogenesis models with right-handed neutrinos with mass greater than the mass scale observed at the LHC by the CMS experiment. The possibility of  generating the required lepton asymmetry with a considerably low value of the $W_R$ mass has been discussed in the context of the resonant leptogenesis scenario \cite{Flanz:1994yx}. In the LRSM, it has been pointed out that successful low-scale leptogenesis with a quasi-degenerate right-handed neutrinos mass spectrum, requires an absolute lower bound of 18 TeV on the $W_R$ mass \cite{Frere:2008ct}. Recently, it was reported that just the right amount of  lepton asymmetry can be produced even for a substantially lower value of the $W_R$ mass ($M_{W_{R}}>3 \tev$) \cite{Dev:2014iva} by considering relatively large Yukawa couplings, which has been updated to $13.1 \tev$ after a more careful analysis in Ref. \cite{Dev:2015vra}. In Refs. \cite{Frere:2008ct, Dev:2014iva}, the lepton number violating gauge scattering processes such as $N_{R} e_{R} \rightarrow \bar{u}_{R} d_{R}$, $N_{R} \bar{u}_{R} \rightarrow e_{R} d_{R}$, $N_{R} d_{R} \rightarrow e_{R} u_{R}$ and $N_{R} N_{R} \rightarrow e_{R} \bar{e}_{R}$ have been analyzed in detail. However, lepton number violating scattering processes with external $W_{R}$ have been ignored on the account of the fact that for a heavy $W_{R}$, there will be a relative suppression of $e^{-m_{W_{R}}/m_{N_{R}}}$ in comparison to the processes with no external $W_{R}$. Now if the $W_{R}$ mass is a few $\tev$'s as suggested by the excess signal at the LHC reported by the CMS experiment then one has to take the latter processes seriously. 
   
In Ref. \cite{Dhuria:2015wwa}, we had first pointed out that the lepton number violating washout processes ($e_{R}^{\pm} e_{R}^{\pm} \rightarrow W_{R}^{\pm} W_{R}^{\pm} $ and $e_{R}^{\pm} W_{R}^{\mp}  \rightarrow   e_{R}^{\mp} W_{R}^{\pm}$)  can be mediated by doubly charged Higgs scalars in the conventional LRSM. Following that, in Ref. \cite{Dev:2015vra} only this channel was considered, and for a particular class of type-I seesaw model with relatively small $M_{N_{R}}$ it was found to have a small contribution, as expected for a large $M_{W_{R}}/M_{N_{R}}$. However the other gauge scattering processes in that scenario are strong enough to give a lower bound of $13.1 \tev$ on the $W_R$ mass. In this Letter, we explore the above lepton number violating scattering processes mediated by both $\Delta^{++}_{R}$ and $N_{R}$ in a much more general context, where we have also taken into account the interference of these channels. The former channel has one gauge vertex and one Yukawa vertex, while for the latter channel both the vertices are gauge vertices, thus are highly unsuppressed compared to the processes involving Yukawa vertices. We find that the lepton number violating scattering process $e_{R}^{\pm} W_{R}^{\mp}  \rightarrow   e_{R}^{\mp} W_{R}^{\pm}$ mediated via both $N_{R}$ and $\Delta^{++}_{R}$  can stay in equilibrium till the electroweak phase transition for a $\tev$ scale $W_{R}$ and wash out the lepton asymmetry \footnote{\label{ft1} Note that the other scattering process is doubly phase space suppressed at a temperature below the $W_{R}$ mass and hence we will not consider it for leptogenesis at $T \lesssim M_{W_{R}}$.}. Thus if one incorporates the above washout process in the Boltzmann equation for lepton number asymmetry, the mentioned lower limit on $M_{W_{R}}$ for successful $\tev$-scale resonant leptogenesis will further go up. In the later variant of LRSM mentioned above the doubly charged Higgs is not there, however, the lepton number violating scattering processes mediated via $N_{R}$ are still present and will wash out the lepton asymmetry.

In the LRSM,  the charged current interaction involving the right-handed neutrino and the right-handed gauge boson is given by
\be
\label{eq:LN}
{\cal L}_{N}= \frac{1}{2 \sqrt{2}}  g_R J_{R\mu} W^{-\mu}_{R}+h.c.
\ee
where $J_{R\mu}= {\bar e_R} \gamma_{\mu} \left(1+\gamma_5 \right) N_{R}$. The Lagrangian for the right-handed Higgs triplet is given by 
\be
\label{eq:KT}
{\cal L}_{\Delta_R} \supset  \left(D_{R \mu} \vec{\Delta}_R\right)^{\dagger}\left(D_{R}^{\mu} \vec{\Delta}_R\right),
\ee
where $\vec{\Delta}_R= \left(\Delta^{++}_R,\Delta^{+}_R,\Delta^{0}_R\right)$ in the spherical basis and the covariant derivative  is defined as $D_{R \mu}= \partial_{\mu}-i g_R \left( T^{j}_{R} A^{j}_{R \mu}\right)- i g' B_{\mu}$. The $A^{j}_{R \mu}$  and $B_{\mu}$ are gauge fields associated with $SU(2)_{R}$ and $U(1)_{B-L}$ groups with the gauge couplings given by $g_R$ and $g'$ respectively.
After spontaneous breaking of the left-right symmetry by giving VEV to  the neutral Higgs field $\Delta^{0}_{R}$ i.e.  $\langle\Delta^{0}_{R}\rangle=\frac{1}{\sqrt 2} v_{R}$, the interaction between the doubly charged Higgs triplet and the gauge boson $W_{R}$  will be given by \cite{Doi:1999tu}
\be
\label{gR}
{\cal L}_{\Delta_R}\supset  
\left(-\frac{v_R}{\sqrt 2} \right)g^{2}_{R} W^{-}_{\mu R} W^{-\mu}_{R} \Delta^{++}_{R} +h.c.
\ee
The Yukawa interaction between the lepton doublet $\psi_{eR}=\left( N_{R},e_{R}\right)^{T}$ and the Higgs triplet $\vec{\Delta}_{R}$ will be given by
\be
\label{Yuk}
{\cal L}_{Y}=  h^{R}_{ee} {\overline{ \left(\psi_{eR}\right)^{c}}} \left( i \tau_2 \vec{\tau} . \vec{\Delta}_{R}\right) \psi_{eR} +h.c. ,
\ee
where $\tau$'s are the Pauli matrices. By giving a VEV to the neutral Higgs triplet field, the Yukawa coupling can be expressed as 
 $h^{R}_{ee}=\frac{ M_{N_R} }{2 v_R}$ 
where $M_{N_R}$ corresponds to mass of the Majorana neutrino ($N_R$).  

The Feynman diagrams of the lepton number violating scattering process induced by the above interactions are shown in Fig. \ref{fig:scattering1}.  
\begin{figure}[h]
	\includegraphics[width=3.4in, height=2.5in]{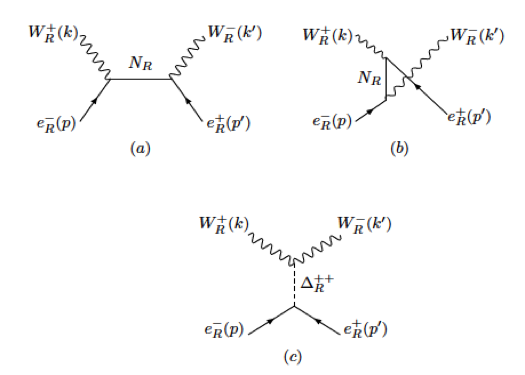}
	\caption{Feynman diagrams for $e^{-}_R W^{+}_R \rightarrow e^{+}_R W^{-}_R$ scattering mediated by $N_R$ and $\Delta^{++}_R$ fields. The Feynman diagrams for $e^{-}_R e^{-}_R \rightarrow W^{-}_R W^{-}_R$ are the same as above with appropriate change in direction of the external lines.}
	\label{fig:scattering1} 
\end{figure}
Utilizing the interactions in Eqs. (\ref{eq:LN})-(\ref{Yuk}), the differential scattering cross section 
for the $e_{R}^{\mp}(p) W_{R}^{\pm}(k)  \rightarrow   e_{R}^{\pm}(p') W_{R}^{\mp}(k')$ 
process is given by \cite{Doi:1999tu}
\be {\label{2.13}}
\frac{ d\sigma^{e_R W_R}_{e_R W_R}}{dt}=\frac{1}{384 \pi M^{4}_{W_R} \left( s - M^{2}_{W_R} \right)^2} \Lambda^{e_R W_R}_{e_R W_R}(s,t,u),
\ee
where 
\be{\label{2.13a}}
\Lambda^{e_R W_R}_{e_R W_R}(s,t,u)=\left. \Lambda^{e_R W_R}_{e_R W_R}(s,t,u)\right|_{N_R}+ \left. \Lambda^{e_R W_R}_{e_R W_R}(s,t,u)\right|_{\Delta^{++}_R}
\ee
and
\bea {\label{2.13b}}
 && \hskip -0.2in
  \left. 
 \Lambda^{e_R W_R}_{e_R W_R}(s,t,u)\right|_{N_R}= g^{4}_{R} \left\{-t \left|M_{N_R}\left(\frac{s}{s- M^{2}_{N_R}}+\frac{u}{u- M^{2}_{N_R}}\right)\right|^2  \right. \nonumber\\
&& \hskip -0.2in \left. -4 M^{2}_{W_R}\left(su-M^{4}_{W_R}\right)\left(s-u\right)^2\left|\frac{M_{N_R}}{\left(s- M^{2}_{N_R}\right)\left(u- M^{2}_{N_R}\right)}\right|^{2} \right. \nonumber\\
&& \hskip -0.2in \left.- 4 M^{4}_{W_R} t \left(\left|\frac{m_{N_R}}{\left(s- M^{2}_{N_R}\right)}\right|^{2}+\left|\frac{M_{N_R}}{\left(u- M^{2}_{N_R}\right)}\right|^{2}\right)\right\},
\eea
\bea {\label{2.13c}}
&&  \hskip -0.2in \left. \Lambda^{e_R W_R}_{e_R W_R}(s,t,u)\right|_{\Delta^{++}_R}
= 4 g^{4}_{R} (-t)  \left\{ \frac{ \left(s+u\right)^2 + 8 M^{4}_{W_R}}{ \left(t- M^{2}_{\Delta_{R}}\right)^2} \left|M_{N_R}\right|^{2}\right. \nonumber\\
&&\left. +\frac{\left(s+u\right)}{t-M^{2}_{\Delta_{R}}}\left|M_{N_R}\right|^{2}\left(\frac{s}{s-M^{2}_{N_R}}+\frac{u}{u-M^{2}_{N_R}}\right) \right. \nonumber\\
&& \left. +\frac{4 M^{4}_{W_R}}{t-M^{2}_{\Delta_R}}\left|M_{N_R}\right|^{2}\left(\frac{1}{s-M^{2}_{N_R}}+\frac{1}{u-M^{2}_{N_R}}\right) \right\},
\eea
where we have neglected any mixing between $W_{L}$ and $W_{R}$. Note that on the r.h.s. of Eq. (\ref{2.13c}) the first term represents the Higgs scalar exchange itself while the last two terms correspond to the interference between the Higgs scalar exchange and the $N_{R}$ exchange mechanisms. The relation between Mandelstem variables  $s=\left(p+k \right)^2, t=\left(p-p' \right)^2$ and $u=\left(p-k'\right)^2$  and scattering angle $\theta$ is given by
\bea {\label{2.14}}
 && \bpm st  \\ su- M^{4}_{W_R}\epm =-\frac{1}{2} \left(s-M^{2}_{W_R}\right)^2 \left(1\mp \cos\theta\right).
\eea

The differential scattering cross section 
for the $e_{R}^{\pm}(p) e_{R}^{\pm}(p') \rightarrow W_{R}^{\pm}(k) W_{R}^{\pm}(k') $
process is given by \cite{Doi:1999tu}
\bea
\label{2.16}
\frac{ d\sigma^{e_R e_R}_{W_RW_R}}{dt}=\frac{1}{512 \pi M^{4}_{W_R} s^2} \Lambda^{e_R e_R}_{W_RW_R}(s,t,u),
\eea
where
\be {\label{2.17}}
  \Lambda^{e_R e_R}_{W_RW_R}(s,t,u)= \left. \Lambda^{e_R e_R}_{W_R W_R}(s,t,u)\right|_{N_R}+ \left. \Lambda^{e_R e_R}_{W_R W_R}(s,t,u)\right|_{\Delta^{++}_R}.
  \ee
  The expressions of  $\Lambda^{e_R e_R}_{W_R W_R}(s,t,u)$ in this case are obtained  by interchanging $s\leftrightarrow t$ in $\Lambda^{e_R W_R}_{e_R W_R}(s,t,u)$: $\Lambda^{e_R e_R}_{W_R W_R}(t,s,u)$= -$\Lambda^{e_R W_R}_{e_R W_R}(s,t,u)$. In this case, the Mandelstem variables $t=\left(p-k \right)^2$ and $u=\left(p-k'\right)^2$ are related to  $s=\left(p+p' \right)^2$ and scattering angle $\theta$  by 
 \bea {\label{2.18}}
 &&  \hskip -0.25in \bpm t  \\ u \epm =-\frac{s}{2} \left(1-\frac{2M^{2}_{W_R}}{s}\right) \left\{1 \mp \sqrt{1-\left(\frac{2 M^{2}_{W_R}}{s-2M^{2}_{W_R}}\right)^2} \cos\theta\right\}. \nonumber\\
\eea
  
During the period  $v_{R} > T > M_{W_{R}}$, both the lepton number violating processes are very fast without any suppression. To get an idea of the effectiveness of these scattering processes in wiping out the lepton asymmetry, we estimate the parameter 
\be
 \label{2.19}
K\equiv \frac{n\langle \sigma \vert v \vert\rangle}{H},
\ee
   for both the processes, where $n$ is the number density of relativistic species and is given by $n=2 \times \frac{3\zeta(3)}{4\pi ^{2}} T^{3}$, $H$ is the Hubble rate given by $H\simeq 1.7 g_{\ast}^{1/2} T^{2}/M_{\rm{Pl}}$, where $g_{\ast} \sim 100$ corresponds to the number of relativistic degrees of freedom, and $\langle \sigma \vert v \vert\rangle$ is the thermally averaged cross section. In order to obtain a rough estimate of $v_{R}$, let us draw an analogy with the Standard Model, where we have $\langle\phi \rangle = \frac{v_L}{\sqrt 2}$ where $v_L=246 \gev$, and $M_{W_{L}} \sim 80 \gev$. Now in the LRSM scenario, where we have 
$\langle\Delta^{0}_{R} \rangle = \frac{v_{R}}{\sqrt 2}$ breaking the left-right symmetry and $M_{W_{R}}=g_R v_R$. Then taking $g_{R} \sim g_{L}$, we have
$\frac{ \langle \phi \rangle}{M_{W_L}} = \frac{\langle\Delta^{0}_{R}\rangle}{M_{W_R}} \approx 3.$

Using the differential cross-section given in Eqs. (\ref{2.13}) and (\ref{2.16}), we plot the behavior of $K$ as a function of temperature in the range $ 3 M_{W_R} > T> M_{W_R}$ for $M_{W_R}=2.1 \tev$ (in the mass range of CMS excess) in Fig. \ref{fig:Kplot}. 
\begin{figure}[h]
\includegraphics[width=3.2in,height=2.1in]{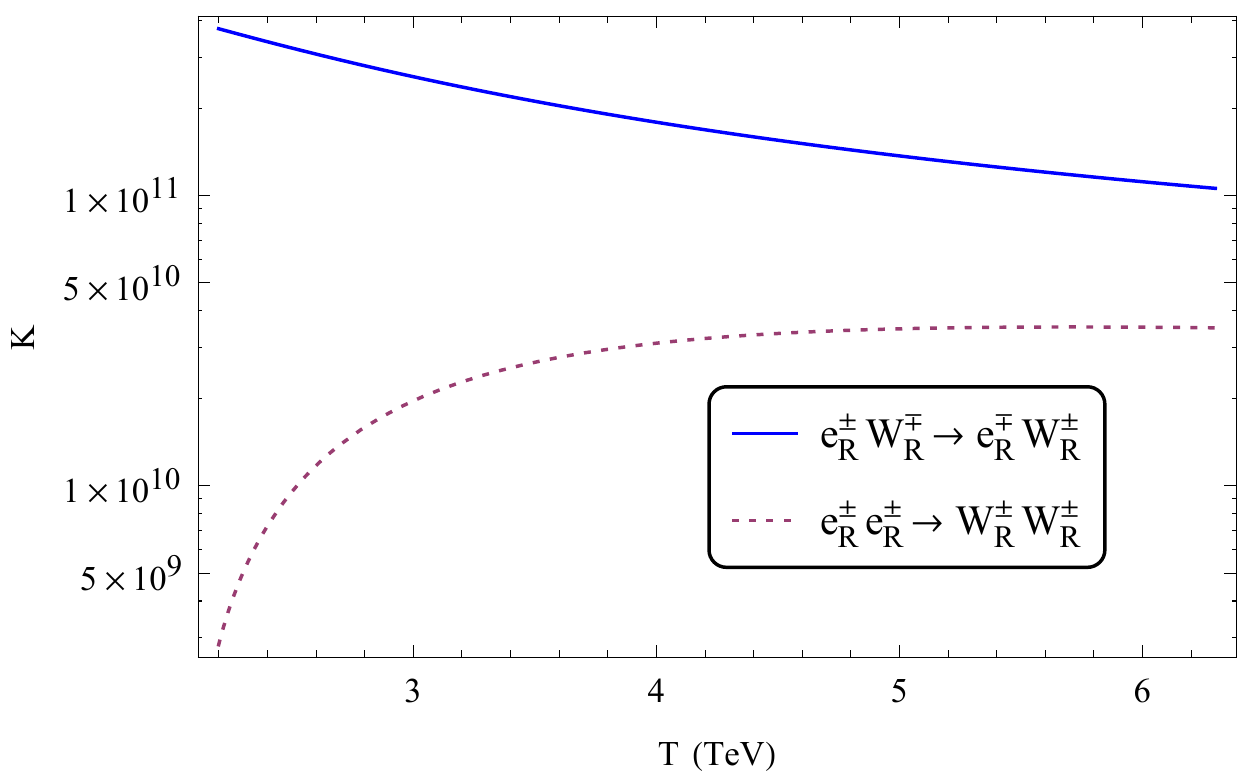}
	\caption{Plot showing $K$  as a function of temperature ($T$) with $M_{W_{R}}=2.1 \tev $ for the scattering processes $e_{R}^{\pm} W_{R}^{\mp}  \rightarrow   e_{R}^{\mp} W_{R}^{\pm}$  and $e_{R}^{\pm} e_{R}^{\pm} \rightarrow W_{R}^{\pm} W_{R}^{\pm} $ (including both $\Delta^{++}_R$ and $N_R$ mediated diagrams) for $v_{R} > T > M_{W_{R}}$. }
	\label{fig:Kplot} 
\end{figure}
The high value of $K$ in Fig. \ref{fig:Kplot} for both the processes implies that  these scattering processes are very fast in washing out lepton asymmetry for  $T \gsim M_{W_{R}}$. In the variant of LRSM with doublet Higgs scalars the scattering processes cannot be mediated via a doubly charged Higgs scalar. However, these lepton number violating scattering processes can still be mediated via heavy neutrinos, which washes out the lepton asymmetry in this scenario for $T \gsim M_{W_{R}}$.\\ 

For $T <  M_{W_{R}}$, the process $e_{R}^{\pm} W_{R}^{\mp}  \rightarrow   e_{R}^{\mp} W_{R}^{\pm}$ is more important \footnotemark[1]. Below we will estimate a lower bound on $T$ till which the latter process stays in equilibrium below $T= M_{W_R}$. The cross section of this process as a function of temperature $T$ can be obtained from Eq. (\ref{2.13}). The scattering rate is given by \footnote{We have ignored any finite temperature effects to simplify the analysis. These effects are small and do not change our conclusions.}
$\Gamma=  \bar{n} \langle \sigma  v_{\rm{rel}} \rangle.$
At a temperature $T< M_{W_{R}}$ the number density $\bar{n}=
g \left(\frac{T M_{W_{R}}}{2\pi}\right)^{3/2}\exp\left(-\frac{M_{W_{R}}}{T}\right)$ 
accounts for the Boltzmann suppression of the scattering rate.
The condition for the scattering process to be in thermal equilibrium is
$\Gamma > H$.
Using $M_{N_{R}} \lsim M_{W_R} $ and $v_{\rm{rel}} = 1$
 \begin{figure}[h] 
\includegraphics[width=3.2in,height=2.1in]{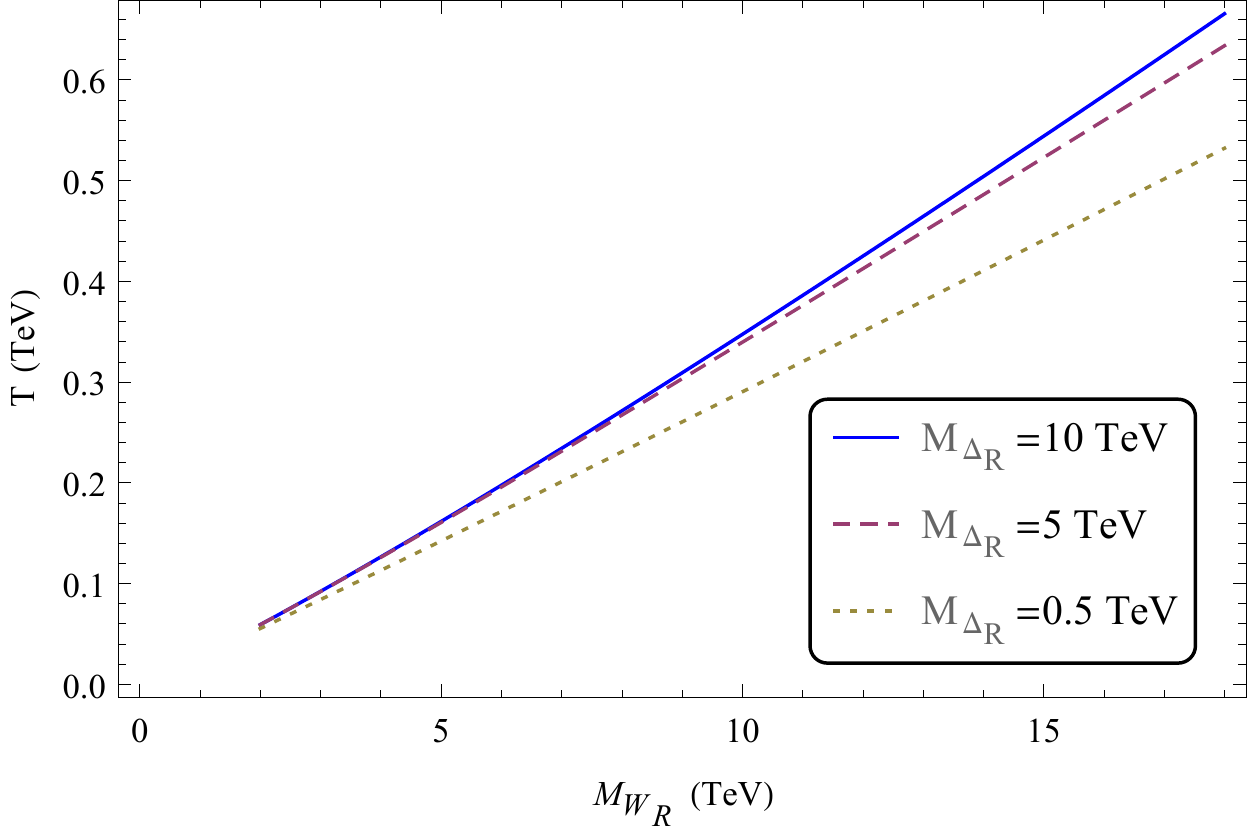}
	 \caption{Plots showing the out-of-equilibrium temperature ($T$) of the scattering process  $e_{R}^{\pm} W_{R}^{\mp}  \rightarrow   e_{R}^{\mp} W_{R}^{\pm}$ (mediated via ${\Delta^{++}_{R}}$ and $N_R$ fields) as a function of $M_{W_{R}}$ for three different values of $M_{\Delta_R}$ and $M_{N_{R}} \sim M_{W_{R}}$.}
 	 \label{fig:graph2} 
 \end{figure}
we plot the temperature till which the scattering process $e_{R}^{\pm} W_{R}^{\mp}  \rightarrow   e_{R}^{\mp} W_{R}^{\pm}$ stays in equilibrium  as a function of the $M_{W_{R}}$  in Fig. \ref{fig:graph2} for three different values of $M_{\Delta_R}$.  We have chosen the lowest value of $M_{\Delta_{R}}$ to be $500 \gev$ in accordance with the recent search limits on the doubly charged Higgs boson mass \cite{Agashe:2014kda}. The plot clearly shows that unless $M_{W_{R}}$ is significantly larger than the $\tev$ scale, the scattering process $e_{R}^{\pm} W_{R}^{\mp}  \rightarrow   e_{R}^{\mp} W_{R}^{\pm}$ will stay in equilibrium till a temperature close to the electroweak phase transition and will continue to wash out the lepton asymmetry till that temperature. In the LRSM scenario with doublet Higgs scalars, the lepton number violating scattering processes mediated only via heavy neutrinos will continue to wash out the asymmetry till the electroweak phase transition, pushing up the lower limit on the $W_{R}$ mass for a successful leptogenesis scenario far beyond the $W_{R}$ signal range reported by the CMS experiment, ruling out the possibility of generating the observed baryon asymmetry from $\tev$ scale resonant leptogenesis as well.

To conclude, for the high-scale leptogenesis scenario ($T \gsim M_{W_{R}}$), in both the variants of the LRSM the lepton number violating scattering processes ($e_{R}^{\pm} e_{R}^{\pm} \rightarrow W_{R}^{\pm} W_{R}^{\pm} $ and $e_{R}^{\pm} W_{R}^{\mp}  \rightarrow   e_{R}^{\mp} W_{R}^{\pm}$) are very efficient in wiping out the lepton asymmetry, while for a $\tev$ scale resonant leptogenesis scenario the latter process will stay in equilibrium till the electroweak phase transition, washing out the lepton asymmetry for $T<M_{W_{R}}$.   Hence we rule out the possibility of successful leptogenesis for $W_{R}^{\pm}$ with mass in the $\tev$ range
\begin{itemize}
	\item in all possible high-scale leptogenesis scenarios for the LRSM variants with (i) triplet Higgs and (ii) doublet Higgs, and
	\item in $\tev$ scale resonant leptogenesis scenarios for LRSM variants with (i) triplet Higgs and (ii) doublet Higgs.
\end{itemize}
Complementing the above results, we have also explored the low-energy subgroups of superstring motivated $E_{6}$ model in recent works.  In one of the supersymmetric low-energy subgroups of the $E_{6}$ (known as the Alternative Left-Right Symmetric Model) one can allow for high-scale leptogenesis, and explain the excess signal at the LHC reported by the CMS experiment from resonant slepton decay. However, the excess signal cannot be explained by right-handed gauge boson decay while allowing leptogenesis, in both supersymmetric and non-supersymmetric low-energy subgroups of superstring motivated $E_{6}$ model \cite{Dhuria:2015hta}. Thus, in light of the above, if the two leptons and two jets excess at the LHC reported by the CMS experiment is indeed due to $W^{\pm}_{R}$ decay, then one needs to resort to a post-electroweak phase transition mechanism to explain the baryon asymmetry of the Universe and in this context, the experiments to observe the neutron-antineutron oscillation \cite{Mohapatra:1980de} or ($B - L$) violating proton decay \cite{Pati:1983zp} will play a crucial role in confirming such possibilities.

 
\end{document}